\documentclass[letterpaper,12pt]{article}

\pdfoutput=1

\usepackage{amsmath}
\usepackage{jheppub,framed}
\usepackage{subcaption}
\usepackage[all]{xy}
\allowdisplaybreaks
\usepackage{mathtools}
\usepackage{slashed}

\def\sgn{\text{sign}}
\def\notInt{\slashed{\cap}}

\usepackage{yfonts}

\DeclareMathOperator{\Gr}{Gr}
\DeclareMathOperator{\Li}{Li}

\def\ket#1{\langle #1 \rangle}

\def\p{\!+\!}
\def\m{\!-\!}

\title{The Sklyanin Bracket and \\ Cluster Adjacency at All Multiplicity}

\author{John~Golden,$^1$}
\author{Andrew~J.~McLeod,$^2$}
\author{Marcus~Spradlin$^3$}
\author{and Anastasia~Volovich$^4$}

\affiliation{\footnotesize $^1$ Leinweber Center for Theoretical Physics and Randall Laboratory of Physics, Department of Physics, University of Michigan Ann Arbor, MI 48109, USA}

\affiliation{\footnotesize $^2$ Niels Bohr International Academy, Blegdamsvej 17, 2100 Copenhagen, Denmark}

\affiliation{$^3$ Department of Physics and Brown Theoretical Physics Center, Brown University, Providence, RI 02912, USA}

\affiliation{$^4$ Department of Physics, Brown University, Providence, RI 02912, USA}

\abstract{We argue that the Sklyanin Poisson bracket on $\Gr(4,n)$ can be used to efficiently test whether an amplitude in planar ${\cal N}=4$ supersymmetric Yang-Mills theory satisfies cluster adjacency. We use this test to show that cluster adjacency is satisfied by all one- and two-loop MHV amplitudes in this theory, once suitably regulated. Using this technique we also demonstrate that cluster adjacency implies the extended Steinmann relations at all particle multiplicities.}

\preprint{
\begin{flushright} LCTP-19-03
\end{flushright}
}

\begin{document}
\maketitle
\flushbottom


\section{Introduction}
\label{Section:Introduction}

Grassmannian cluster algebras have been found to play a remarkable role in the amplitudes of planar ${\cal N}=4$ supersymmetric Yang-Mills (sYM) theory~\cite{Golden:2013xva}. Many aspects of these amplitudes exhibit cluster-algebraic structure---from their integrands~\cite{ArkaniHamed:2012nw} to their symbol alphabets~\cite{,Golden:2013lha,Golden:2014pua} and cobrackets~\cite{Golden:2014xqa,Golden:2014xqf,Harrington:2015bdt,Golden:2018gtk}---despite the fact that the physics underlying most of these features remains obscure. Recently there has been renewed interest in this connection due to a conjecture that certain amplitudes respect a `cluster adjacency' principle~\cite{Drummond:2017ssj,Drummond:2018dfd}. This principle states that cluster coordinates only appear in adjacent symbol entries of the $n$-point amplitude if they also appear together in a cluster of $\Gr(4,n)$. This surprising property has been shown to hold in all known six- and seven-point amplitudes~\cite{CaronHuot:2011ky,Dixon:2014iba,Drummond:2014ffa,Dixon:2015iva,Caron-Huot:2016owq,Dixon:2016nkn,Dixon:2016apl,Drummond:2018dfd,Drummond:2018caf}, as well as some eight- and nine-point amplitudes~\cite{Drummond:2017ssj,Drummond:2018dfd}. In six-particle kinematics it is also consistent with an all-orders analysis of the Landau equations~\cite{Prlina:2018ukf}, and is obeyed by certain classes of Feynman integrals to all loop orders~\cite{Caron-Huot:2018dsv}.

In this paper we show that cluster adjacency holds for suitably normalized one- and two-loop MHV amplitudes at arbitrary $n$. While this class of amplitudes has been known for many years~\cite{Bern:1994zx,CaronHuot:2011kk}, checking cluster adjacency for $n>7$ appears to be nontrivial due to the nature of the $\Gr(4,n>7)$ cluster algebras---both the number of clusters and the number of cluster coordinates in each of these algebras is infinite. This makes it hard to rule out the possibility that any given pair of cluster coordinates appears together in a cluster, as no closed-form expression for this infinite set of cluster coordinates (or the clusters into which they combine) is known in these cases. Even so, this complication can in some cases be circumvented, for instance by identifying finite subalgebras of $\Gr(4,n>7)$~\cite{Drummond:2018dfd,inProgressBGM} (as can be done, for instance, in the multi-Regge limit~\cite{DelDuca:2016lad}). Here we instead utilize the Sklyanin Poisson bracket~\cite{Sklyanin:1982tf,GSV}, which can be computed for any pair of cluster coordinates. It is conjectured\footnote{We in particular thank M.~Gekhtman for correspondence and discussions on this topic.} that this bracket is (half-)integer valued (in a sense that will be made precise below) when these coordinates appear together in at least one cluster.

Loop-level amplitudes such as the ones we consider suffer from infrared divergences that need to be regulated. Due to the freedom to shift the normalization of these amplitudes by any finite function, there is a danger that cluster adjacency may be spoiled in some regularization schemes. Indeed, it is known that the BDS remainder function (defined in~\cite{Bern:2005iz}) does not satisfy cluster adjacency at six or seven points, while the BDS-like normalized amplitude (which was first considered at strong coupling~\cite{Alday:2009dv,Yang:2010as}) does. Even before the discovery of cluster adjacency, the latter normalization had received increased attention in the literature because the BDS-like normalization preserves the Steinmann relations~\cite{Steinmann,Steinmann2,Cahill:1973qp}, which constrain the iterated discontinuity structure of these amplitudes (and thereby also their adjacent symbol letters)~\cite{Caron-Huot:2016owq,Dixon:2016nkn}. While the BDS-like amplitude can only be defined when $n$ is not a multiple of four, there exist normalization schemes in which the Steinmann relations can be preserved for arbitrary $n$~\cite{Golden:2018gtk}, and in which cluster adjacency has been checked to hold for eight points~\cite{cluster_subalgebras_ii} at two loops. We define one such `minimal' normalization scheme below, and show that one- and two-loop MHV amplitudes satisfy cluster adjacency in this scheme for all $n$, and in the BDS-like scheme for all $n$ in which it can be defined.

In six-particle kinematics, cluster adjacency and the extended Steinmann relations become equivalent once supplemented with the requirement of single-valuedness in the Euclidean region~\cite{cosmic_galois_paper}. However, this has been shown only by the explicit construction of the full space of appropriate polylogarithmic functions to high weight, a procedure that is computationally infeasible at higher $n$. Using the Sklyanin bracket, we clarify the relation between cluster adjacency and the extended Steinmann relations by showing that any polylogarithmic function satisfying the former automatically satisfies the latter. It is not yet known whether the two conditions are equivalent in general, although cluster adjacency na\"ively imposes a larger set of constraints.

This paper is organized as follows. In section~\ref{sec:clusterreview} we review a few necessary facts about cluster algebras with a particular focus on the Sklyanin bracket and its applicability for testing cluster adjacency. In section~\ref{sec:mhv_amplitudes} we review the problem of constructing suitably regulated amplitudes, recall the definition of the BDS-like amplitude for ${4\! \not\vert \  n}$, and introduce a new minimally subtracted amplitude for all $n$. Sections~\ref{sec:camhv} and~\ref{sec:cluster_adjacency_implies_steinmann} contain the main results of this paper. In the former we show that all one- and two-loop MHV amplitudes satisfy cluster adjacency, and in the latter we demonstrate in general that the extended Steinmann relations follow from cluster adjacency.

Finally, in an ancillary {\sc Mathematica} notebook we include code for generating the $n$-particle one-loop amplitude in both BDS-like and minimally-regulated schemes, and for computing the Sklyanin bracket between all pairs of adjacent coordinates in any symbol. We explicitly provide the analogous two-loop symbols at low $n$, but more generally provide code to convert the output of the notebook accompanying~\cite{CaronHuot:2011ky} into either of the regularization schemes considered here.

\section{Cluster Algebras and Poisson Brackets}
\label{sec:clusterreview}

There are several excellent reviews on cluster algebras and their appearance in planar $\mathcal{N}=4$ sYM (see for example~\cite{ArkaniHamed:2012nw,Golden:2013xva,Paulos:2014dja,Vergu:2015svm,Golden:2018gtk}); here we review only the bare minimum needed for our purposes. The two main takeaways from this section are (i) how to compute a Poisson bracket between two cluster $\mathcal{A}$-coordinates, and (ii) that this provides an efficient test for determining whether there exists a cluster containing two given cluster coordinates $a_1$ and $a_2$; namely, this is the case if and only if $\{ \log a_1, \log a_2 \} \in \frac{1}{2} \mathbb{Z}$.

\subsection{Cluster Coordinates}
\label{sec:first}

A \emph{cluster algebra} of rank $r$ is a commutative algebra whose generators, called \emph{cluster} $\mathcal{A}$-\emph{coordinates}, are assembled into sets of $r$ elements called \emph{clusters}, with the $\mathcal{A}$-coordinates in any cluster related to those in any other cluster by an operation called \emph{mutation}. Some cluster algebras have the property that each cluster can be associated with a \emph{quiver} (an oriented graph with no one- or two-cycles) formed out of $r$ vertices, in which each vertex is labeled by one of the $\mathcal{A}$-coordinates. The structure of each quiver is encoded in its associated $r \times r$ \emph{exchange matrix}
\begin{align}
	\label{eqn:bijdef}
b_{ij} = ( \#~{\rm of~arrows~} i \to j) - ( \#~{\rm of~arrows~} j \to i )\,.
\end{align}
Note that $b_{ij}$ is manifestly antisymmetric and integer-valued. This matrix dictates how a cluster changes under mutation.

\begin{figure}[t]
\centering
\begin{subfigure}[b]{0.45\textwidth}
\begin{align*}
\begin{gathered}
\begin{xy} 0;<-.5pt,0pt>:<0pt,-.5pt>::
         (-100,0) *+{\framebox[7ex]{$\ket{1234}$}} ="0",
        (0,0) *+{\ket{2346}} ="1",
        (100,0) *+{\framebox[7ex]{$\ket{2345}$}} ="2",
        (0,75) *+{\ket{1346}} ="3",
        (100,75) *+{\framebox[7ex]{$\ket{3456}$}} ="4",
        (0,150) *+{\ket{1246}} ="5",
        (100,150) *+{\framebox[7ex]{$\ket{1456}$}} ="6",
        (0,225) *+{\framebox[7ex]{$\ket{1236}$}} ="7",
        (100,225) *+{\framebox[7ex]{$\ket{1256}$}} ="8",
        "0", {\ar"1"},
        "1", {\ar"2"},
        "3", {\ar"4"},
        "5", {\ar"6"},
        "1", {\ar"3"},
        "3", {\ar"5"},
        "5", {\ar"7"},
        "4", {\ar"1"},
        "6", {\ar"3"},
        "8", {\ar"5"},
\end{xy}
\end{gathered}
\end{align*}
\caption{} \label{fig:g46-a-seed}
\end{subfigure}
\begin{subfigure}[b]{0.25\textwidth}
\begin{align*}
\begin{gathered}
\begin{xy} 0;<-.5pt,0pt>:<0pt,-.5pt>::
        (0,0) *+\txt{\fontsize{16pt}{16pt} $\frac{\ket{1234}\ket{3456}}{\ket{2345}\ket{1346}}$} ="1",
        (0,100) *+\txt{\fontsize{16pt}{16pt} $\frac{\ket{2346}\ket{1456}}{\ket{3456}\ket{1246}}$} ="3",
        (0,200) *+\txt{\fontsize{16pt}{16pt} $\frac{\ket{1346}\ket{1256}}{\ket{1456}\ket{1236}}$} ="5",
        "1", {\ar"3"},
        "3", {\ar"5"},
\end{xy}
\end{gathered}
\end{align*}
\caption{} \label{fig:g46-x-seed}
\end{subfigure}
\caption{(a) One of the 14 quivers for the $\Gr(4,6)$ cluster algebra, with each vertex labeled by its associated $\mathcal{A}$-coordinate and the frozen coordinates in boxes. (b) The same quiver with frozen vertices omitted and with each mutable vertex labeled by its associated $\mathcal{X}$-coordinate.}
\label{fig:g46-seed}
\end{figure}
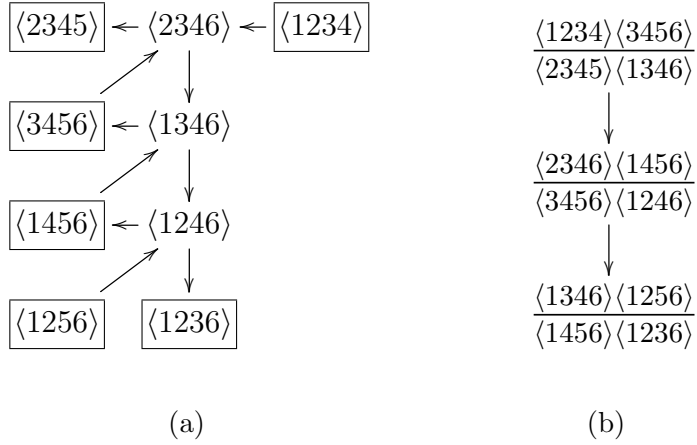

The cluster algebra relevant to scattering amplitudes in planar sYM theory, $\Gr(4,n)$, is moreover of \emph{geometric type}, which means that a specific subset of the vertices in each quiver are \emph{frozen}, and their associated \emph{frozen coordinates} do not change under mutation. Correspondingly, each frozen coordinate is a member of every cluster (and is not counted towards the rank of the algebra). The frozen coordinates of the $\Gr(4,n)$ cluster algebra are the cyclic Pl\"ucker coordinates $\langle 1234\rangle, \langle 2345\rangle, \ldots, \langle 123n \rangle$, which correspond (as will be reviewed below) to two-particle Mandelstam invariants. Each cluster contains, in addition to these frozen coordinates, $3(n-5)$ \emph{mutable} (non-frozen) $\mathcal{A}$-coordinates. As an example, Figure~\ref{fig:g46-a-seed} shows one of the 14 quivers for the $\Gr(4,6)$ cluster algebra.

The $\Gr(4,n)$ cluster algebra has a closely related \emph{cluster Poisson variety}~\cite{FG03b}, in which each mutable vertex of every quiver gets labelled by a \emph{cluster} $\mathcal{X}$-\emph{coordinate}. One such assignment of $\mathcal{X}$-coordinates\footnote{For other ways of assigning $\mathcal{X}$-coordinates to these vertices, see for example~\cite{ArkaniHamed:2012nw,positroidTwists,Bourjaily:2018aeq}.} can be formed directly out of the $\mathcal{A}$-coordinates of these clusters via
\begin{align} \label{eq:a_to_x}
x_\alpha = \prod_i a_i^{b_{j\alpha}}\,.
\end{align}
(Here and below, Greek indices are taken to range only over mutable vertices, while Latin indices range over all vertices.) Figure~\ref{fig:g46-x-seed} shows the $\mathcal{X}$-coordinate quiver that arises from applying the map~\eqref{eq:a_to_x} to the $\mathcal{A}$-coordinate quiver in Figure~\ref{fig:g46-a-seed}.

There exists a natural Poisson bracket on $\Gr(4,n)$, which can be computed between two $\mathcal{X}$-coordinates $x'_\rho$ and $x'_\sigma$ as
\begin{align} \label{eq:poisson}
\{ \log x^\prime_\rho, \log x^\prime_\sigma \} \equiv \sum_{\alpha, \beta = 1}^r \left( \frac{\partial \log x^\prime_\rho}{\partial \log x_\alpha} \right) b_{\alpha \beta} \left( \frac{\partial \log x^\prime_\sigma}{\partial \log x_\beta} \right),
\end{align}
where the variables ${x}_\alpha$ correspond to a complete set of $\mathcal{X}$-coordinates drawn from an arbitrary reference cluster. In general, $x^\prime_\rho$ and $x^\prime_\sigma$ will be complicated rational functions of the reference $\mathcal{X}$-coordinates, in which case the Poisson bracket will also evaluate to a complicated rational function of these variables. However, if there exists a cluster containing both $x^\prime_\rho$ and $x^\prime_\sigma$, the bracket~\eqref{eq:poisson} must compute the relevant entry $b'_{\rho\sigma}$ of the exchange matrix describing this cluster (as can be seen by choosing this cluster to be the reference cluster). This implies, in particular, that the result will be an integer. More importantly, the converse of this statement is also expected to hold---namely, if $\{ \log x^\prime_\rho, \log x^\prime_\sigma \} \in \mathbb{Z}$, then there exists a cluster containing both $x^\prime_\rho$ and $x^\prime_\sigma$.\footnote{This statement is somewhat too strong; the correct statement is that if $\{ \log x^\prime_\rho, \log x^\prime_\sigma \} \in \mathbb{Z}$, then there exists a cluster containing both elements of one of the four pairs $(x^\prime_\rho, x^\prime_\sigma)$,  $(x^\prime_\rho, 1/x^\prime_\sigma)$, $(1/x^\prime_\rho, 1/x^\prime_\sigma)$, or $(1/x^\prime_\rho, x^\prime_\sigma)$. The bracket defined in~\eqref{eq:poisson} only changes sign under taking $x \to 1/x$ so it evaluates to an integer in all four cases and this test cannot be used to diagnose which of the four possibilities is correct. Fortunately this will not matter for us in what follows, since symbol-level cluster adjacency is also insensitive to the distinction between $x$ and $1/x$.} For finite algebras ($n < 8$) it can be checked that this is true by explicitly enumerating all clusters and their associated coordinates. For the infinite algebras ($n > 7$) this expectation remains, to our knowledge, conjectural.

\subsection{The Sklyanin Bracket on $\mathcal{X}$-Coordinates}
\label{sec:sklyanin}

In~\eqref{eqn:bijdef} and~\eqref{eq:poisson} we saw that the Poisson bracket is encoded in the combinatorial structure of a quiver. As we now review, it can alternatively be computed using the famous~\cite{Sklyanin:1982tf,GSV} Sklyanin Poisson bracket on $\Gr(4,n)$. (The reason this works will be clarified in the following section.) Given a $4 \times n$ matrix, which represents a point in $\Gr(4,n)$ and whose columns can be interpreted as momentum twistors $Z_a$ specifying the kinematics of a massless $n$-particle scattering event,
\begin{align}
Z = \left(\begin{array}{ccc}
Z^1_{\ 1} & \ldots & Z^1_{\ n} \\
Z^2_{\ 1} & \ldots & Z^2_{\ n} \\
Z^3_{\ 1} & \ldots & Z^3_{\ n} \\
Z^4_{\ 1} & \ldots & Z^4_{\ n} \end{array}\right),
\end{align}
we can generically put it into the form
\begin{align} \label{eq:gauge_fixed_Z}
Z'=\left(
\begin{array}{ccccccc}
 1 & 0 & 0 & 0 & y_{\ 5}^1 & \ldots  & y_{\ n}^1 \\
 0 & 1 & 0 & 0 & y_{\ 5}^2 & \ldots  & y_{\ n}^2 \\
 0 & 0 & 1 & 0 & y_{\ 5}^3 & \ldots  & y_{\ n}^3 \\
 0 & 0 & 0 & 1 & y_{\ 5}^4 & \ldots  & y_{\ n}^4 \\
\end{array}
\right)
\end{align}
by a suitable $GL(4)$ transformation. The nontrivial elements of the reduced matrix are clearly given by $y_{\ a}^I=(-1)^I \ket{\{1,2,3,4\}\setminus\{I\},a}/\ket{1\,2\,3\,4}$ for $a \in \{5,\ldots,n\}$. The Sklyanin bracket on the $y_{\ a}^I$ variables is then given by:
\begin{framed}
\vspace{-0.4cm}
\begin{align}
\label{eqn:sklyanin}
	\{ y_{\ a}^I, y_{\ b}^J \} = \frac{1}{2} (\sgn(J - I) - \sgn(b - a)) y_{\ a}^J y_{\ b}^I\,.
\end{align}
\vspace{-0.65cm}
\end{framed}
\noindent
This extends to arbitrary functions on $\Gr(4,n)$ in the obvious way, via
\begin{align}
\label{eq:poisson_br}
	\{ f(y), g(y) \} = \sum_{a,b=5}^n\sum_{I,J=1}^4 \frac{\partial f}{\partial y_{\ a}^I}
        \frac{\partial g}{\partial y_{\ b}^J} \{ y_{\ a}^I,  y_{\ b}^J \}\,.
\end{align}
For example, a straightforward if somewhat tedious calculation using~\eqref{eqn:sklyanin} and~\eqref{eq:poisson_br} reveals that
\begin{multline}
        \left\{ \log \frac{\ket{1234}\ket{3456}}{\ket{2345}\ket{1346}}, \log
        \frac{\ket{1346}\ket{1256}}{\ket{1456}\ket{1236}} \right\}\\
	=
\left\{ \log \frac{(1) (y_{\ 5}^1 y_{\ 6}^2{-}y_{\ 6}^1 y_{\ 5}^2)}{(y_{\ 5}^1)(y_{\ 6}^2)}, \log \frac{(y_{\ 6}^2) (y_{\ 5}^3 y_{\ 6}^4{-}y_{\ 6}^3 y_{\ 5}^4)}{(y_{\ 5}^2 y_{\ 6}^3 - y_{\ 6}^2 y_{\ 5}^3)( y_{\ 6}^4)} \right\} = 0\,,
\end{multline}
and similarly
\begin{align}
	&\left\{ \log \frac{\ket{1234}\ket{3456}}{\ket{2345}\ket{1346}},
\log \frac{\ket{2346}\ket{1456}}{\ket{3456}\ket{1246}} \right\}
= 1 \, ,\\
&\left\{ \log \frac{\ket{2346}\ket{1456}}{\ket{3456}\ket{1246}},
\log \frac{\ket{1346}\ket{1256}}{\ket{1456}\ket{1236}} \right\}
	= 1\,.
\end{align}
Both examples are in accordance with the structure of arrows in Figure~\ref{fig:g46-x-seed}. In contrast, one can check for example that
\begin{align}
	\left\{\log \frac{\ket{1234}\ket{3456}}{\ket{2345}\ket{1346}},
	\log \frac{\ket{1236}\ket{1245}}{\ket{1234}\ket{1256}}\right\} \notin \mathbb{Z} \, ,
\end{align}
namely that it evaluates to a nontrivial function on $\Gr(4,n)$, in accordance with the fact that there is no cluster containing both of these $\mathcal{X}$-coordinates.

\subsection{The Sklyanin Bracket on $\mathcal{A}$-Coordinates}
\label{sec:sklyanin-a-test}

For the purpose of testing the cluster adjacency of polylogarithmic scattering amplitudes in perturbative sYM theory, it would be beneficial to have an efficient test for checking the existence of a cluster containing any two given $\mathcal{A}$-coordinates. To that end, we discuss in this section the construction of suitable Poisson brackets on $\mathcal{A}$-coordinates.

Consider an algebra with $f$ frozen vertices and $r$ mutable vertices (in $\Gr(4,n)$, $f = n$ and $r=3(n-5)$). Let $t=r+f$, and let $B$ be the $t \times t$ exchange matrix in some cluster with entries $b_{ij}$ as defined in section~\ref{sec:first}. Then any skew-symmetric $t \times t$ matrix $\Omega$ satisfying
\begin{align}
\label{eqn:omegaB}
        \Omega B^{\rm T} = \left( \begin{matrix}
                {1}_{r \times r} & \ast_{r \times f} \\
                0_{f \times r} & \ast_{f \times f}
        \end{matrix}\right),
\end{align}
where $\ast_{r \times f}$ and $\ast_{f \times f}$ are allowed to have any entries, provides an appropriate Poisson bracket on $\mathcal{A}$-coordinates. That is, if we declare that the $\mathcal{A}$-coordinates in some cluster satisfy
\begin{align}
\{ \log a_i, \log a_j \} = \Omega_{ij}\,,
\end{align}
then the $\mathcal{A}$-coordinates $a_i'$ in any other cluster will satisfy
\begin{align}
\{ \log a_i', \log a_j' \} = \Omega_{ij}' \, ,
\end{align}
where $\Omega_{ij}'$ is the mutated avatar of $\Omega_{ij}$. The precise mutation relation for $\Omega$ can be worked out in principle, but is not very illuminating.

Of course there is considerable ambiguity in choosing $\Omega$---the relation~\eqref{eqn:omegaB} can be understood as the requirement that this ambiguity must drop out when computing the Poisson bracket of any function of $\mathcal{X}$-coordinates, which is canonically defined as reviewed in the previous section. This ambiguity also jibes with the fact that a pair of $\mathcal{A}$-coordinates can appear together in many different clusters, but the (signed) number of arrows $b_{ij}$ connecting their associated vertices will in general be different in different clusters. In contrast, if two $\mathcal{X}$-coordinates satisfy $\{ \log x_i, \log x_j \} = b_{ij} \in \mathbb{Z}$, then they are always connected by precisely $b_{ij}$ appropriately-directed arrows whenever they appear in a cluster together.

Now, finally, we come to the point: if we choose $\Omega$ to be the Sklyanin bracket defined in~\eqref{eqn:sklyanin}, then~\eqref{eqn:omegaB} is satisfied. This is in fact the reason \emph{why} one can use the Sklyanin bracket to compute the Poisson bracket of two $\mathcal{X}$-coordinates, and it can be used just as well on $\mathcal{A}$-coordinates. It is illustrative to consider the cluster shown in Figure~\ref{fig:g46-seed} as an example. If we order the $\mathcal{A}$-coordinates as
\begin{align*}
\{ a_1, \ldots, a_9 \} =
\{ \langle 2346 \rangle,\langle 1346\rangle,
\langle 1246\rangle,
\langle 1234\rangle, \langle 2345 \rangle,
\langle 3456 \rangle, \langle 1456 \rangle,
\langle 1256 \rangle, \langle 1236 \rangle\}\,,
\end{align*}
then the exchange matrix $B$ can be read off from Figure~\ref{fig:g46-a-seed}, and taking $\Omega$ to be the Sklyanin bracket $\Omega_{ij} = \{ \log a_i, \log a_j \}$ computed using the definition~\eqref{eqn:sklyanin}, we find
\begin{align*}
B = \left(
\begin{array}{rrrrrrrrr}
0 & 1 & 0 & -1 & \ 1 & -1 & 0 & \ 0 & 0 \\
-1 & 0 & 1 & 0 & 0 & 1 & -1 & 0 & 0 \\
0 & -1 & 0 & 0 & 0 & 0 & -1 & 1 & -1 \\
1 & 0 & 0 & 0 & 0 & 0 & 0 & 0 & 0 \\
-1 & 0 & 0 & 0 & 0 & 0 & 0 & 0 & 0 \\
1 & -1 & 0 & 0 & 0 & 0 & 0 & 0 & 0 \\
0 & 1 & -1 & 0 & 0 & 0 & 0 & 0 & 0 \\
0 & 0 & 1 & 0 & 0 & 0 & 0 & 0 & 0 \\
0 & 0 & -1 & 0 & 0 & 0 & 0 & 0 & 0
\end{array}
\right), \quad
\Omega = \frac{1}{2} \left(
\begin{array}{rrrrrrrrr}
0& 1& 1& \ 0& 1& 0& 1&\ 1& \ 1 \\
-1& 0& 1& 0& 0& 0& 0& 1& 1 \\
-1& -1& 0& 0& 0& -1& 0& 0& 1 \\
0& 0& 0& 0& 0& 0& 0& 0& 0 \\
-1& 0& 0& 0& 0& 0& 1& 1& 0 \\
0& 0& 1& 0& 0& 0& 1& 2& 1 \\
-1& 0& 0& 0& -1& -1& 0& 1& 1 \\
-1& -1& 0& 0& -1& -2& -1& 0& 0 \\
-1& -1& -1& 0& 0& -1& -1& 0& 0
\end{array}
\right) .
\end{align*}
It is easily verified that~\eqref{eqn:omegaB} is satisfied.

Following the terminology introduced in~\cite{Drummond:2017ssj}, we say that two cluster $\mathcal{A}$-coordinates are \emph{cluster adjacent} if there exists a cluster containing both of them. We conclude this section with the main conjecture, which (like the one at the end of the previous section) is widely believed to be true, although we are not aware of anywhere it is explicitly written down:

\begin{framed}
\begin{center}
\vspace{-.4cm}
Two $\mathcal{A}$-coordinates $a_1, a_2$ exist in a cluster together, and are thus \\ cluster adjacent, if and only if their Sklyanin bracket $\{\log a_1, \log a_2 \} \in \frac{1}{2} \mathbb{Z}$.
\vspace{-.4cm}
\end{center}
\end{framed}

\section{MHV Amplitudes} \label{sec:mhv_amplitudes}

One goal of this paper is to demonstrate that one- and two-loop MHV amplitudes in sYM theory satisfy cluster adjacency. Because of the necessity to regulate infrared divergences beyond tree level, there is some scheme dependence in how to define these amplitudes, and only certain choices yield amplitudes that have a chance to satisfy cluster adjacency (as will be seen very concretely in section~\ref{sec:cluster_adjacency_implies_steinmann}). In particular, the choice of regulator should not spoil the (iterated) discontinuity structure of the amplitude. In this section we briefly review these issues and present new, explicit formulas for two appropriate regularizations: (i) the BDS-like normalized MHV amplitude $\mathcal{E}_n$~\cite{Alday:2009dv,Yang:2010as}, which can be defined whenever $n$ is not a multiple of four, and (ii) a `minimally' normalized MHV amplitude ${\cal{E}}_n^{\rm min}$ that can be defined for any $n$, at the cost of breaking dual conformal symmetry.

\subsection{The BDS-Like Normalized MHV Amplitude}

We begin with the so-called BDS-like normalization, since it is already familiar in the literature~\cite{Alday:2009dv,Yang:2010as}. We denote by ${A}_n$ the $n$-point MHV amplitude in planar sYM theory, with gauge group $SU(N_c)$ and coupling constant $g_{\rm YM}$, in which the tree-level amplitude has been scaled out. We evaluate this amplitude in $4 - 2 \epsilon$ dimensions to regulate infrared divergences. To eliminate these divergences, one must divide ${A}_n$ by some normalization factor with the same divergences, and then take the limit $\epsilon \to 0$. Of course, in carrying out this step there is always an ambiguity in the finite part of the normalizing factor.

The BDS-like normalization is defined, whenever $n$ is not a multiple of four, by the function
\begin{align} \label{eq:BDS_like_def}
{A}_n^{\text{BDS-like}} \equiv \text{exp} \left[ \sum_{L=1}^{\infty} g^{2L} \left( \frac{f^{(L)}(\epsilon) }{2} {A}_n^{(1)}(L\epsilon) + C^{(L)} \right) \right]
\exp\left(\frac{\Gamma_{\text{cusp}}}{4} Y_n\right).
\end{align}
Here $g^2 = \frac{g_{\rm YM}^2 N_c}{16 \pi^2}$ is the 't Hooft coupling, the superscript $(L)$ on any function denotes the ${\cal O}(g^{2 L})$ term in its Taylor series expansion around $g = 0$, and $C^{(L)}$ and $f(\epsilon)$ are transcendental constants. In particular, $f(\epsilon) = \frac{1}{2} \Gamma_{\text{cusp}} + {\cal O}(\epsilon)$, where
\begin{align}
\Gamma_{\text{cusp}} = 4 g^2 - 8 \zeta(2) g^4 + {\cal O}(g^6)
\end{align}
is the planar cusp anomalous dimension~\cite{Beisert:2006ez}. The first exponential in~\eqref{eq:BDS_like_def} is the BDS ansatz, which is known to capture all infrared divergences of ${A}_n$~\cite{Bern:2005iz} in terms of those of the one-loop amplitude ${A}_n^{(1)}$ (first computed in~\cite{Bern:1994zx}), and the second exponential is a certain specific choice for the abovementioned finite ambiguity. The function $Y_n$ is a weight-two polylogarithm, and can be defined in terms of $A_{\text{BDS-like}}$ as given on page 57 of~\cite{Alday:2010vh} and twice the function defined in Eq.~(4.57) of~\cite{Bern:2005iz}, which we denote by $F_n$. (That is, we define $F_n$ here to be two times the expression given this name in~\cite{Bern:2005iz}.) In terms of these quantities, we have
\begin{align}\label{eqn:ydef}
Y_n = - F_n - 4A_{\text{BDS-like}}+ \frac{n \pi^2}{4}\,.
\end{align}
Finally, the BDS-like normalized amplitude $\mathcal{E}_n$ is defined to be
\begin{align} \label{eq:BDS_like_normalized_def}
{\cal E}_n \equiv \left.\frac{{A}_n}{{A}_n^{\text{BDS-like}}}
\right|_{\epsilon=0} = \exp \left(
R_n - \frac{\Gamma_\text{cusp}}{4} Y_n\right)
\qquad
4\! \not\vert \  n\,,
\end{align}
where we have also indicated its relation to the remainder function $R_n$~\cite{Bern:2008ap,Drummond:2008aq}.

The choice of $Y_n$ is motivated by the fact that it is the unique dual conformally invariant function one can append to the BDS ansatz, in the form shown in~\eqref{eq:BDS_like_def}, that renders ${A}_n^{\text{BDS-like}}$ a function of two-particle Mandelstam invariants alone. (No such function exists when $n$ is a multiple of four.) The significance of this lies in the fact that Mandelstam invariants translate to the ratio of Pl\"ucker coordinates
\begin{align}
\label{eqn:mandelstam_translation}
s_{i,i+1,\dots,j} = (p_{i} + p_{i+1} + \dots + p_{j})^2 =
\frac{\ket{ i\m 1 \,i \, j \, j\p 1}}{
\langle i\m 1\, i \, I \rangle \langle j \, j \p 1 \, I\rangle} \, ,
\end{align}
where $I$ is the infinity twistor. Thus, two-particle Mandelstam invariants (up to factors involving the infinity twistor, which cancel out of any dual conformally invariant quantity) only involve frozen coordinates (the cyclic Pl\"uckers). This implies one can safely divide the amplitude ${A}_n$ by ${A}_n^{\text{BDS-like}}$ without spoiling cluster adjacency.

The BDS-like normalization is advantageous for another, related reason. The fact that ${A}_n^{\text{BDS-like}}$ depends only on two-particle invariants means that $\mathcal{E}_n$ satisfies the same Steinmann relations~\cite{Steinmann,Steinmann2,Cahill:1973qp} with respect to multi-particle Mandelstam invariants as the unregulated amplitude ${A}_n$, namely
\begin{align} \label{eq:steinmann}
\begin{rcases}
\text{Disc}_{s_{j,\dots,j+p+q}}\left[\text{Disc}_{s_{i,\dots,i+p}} \big({A}_n \big) \right] &\!\!= \ 0, \hspace{.3cm} \\
\text{Disc}_{s_{i,\dots,i+p}}\left[\text{Disc}_{s_{j,\dots,j+p+q}} \big({A}_n \big) \right] &\!\!= \ 0,
\end{rcases} \quad
\begin{gathered} 0 < j\!-\!i \leq p \text{\ \ or\ \ } q < i\!-\!j  \leq p\!+\!q\,. \end{gathered}
\end{align}
In the case of planar ${\cal N}=4$ sYM, where several infinite classes of amplitudes are believed to be polylogarithms of uniform transcendental weight, this directly translates into a condition on the \emph{symbol} of these amplitudes~\cite{Caron-Huot:2016owq,Dixon:2016nkn}.\footnote{For an introduction to how symbols~\cite{Goncharov:2010jf} encode the discontinuity structure of polylogarithms, see for instance~\cite{Duhr:2014woa}.} Specifically, the Steinmann relations forbid the $\mathcal{A}$-coordinates $\ket{i\m 1\  i \ i \p p \ i \p p \p 1}$ and $\ket{j\m 1\ j\ j\p p \p q \ j\p p \p q \p 1}$ from appearing in the first two entries of the symbol when either of the inequalities in~\eqref{eq:steinmann} is satisfied. While the Steinmann relations were originally conceived as constraints on the first two discontinuities of any amplitude, they have recently been observed to hold to all depth in the symbol of planar sYM amplitudes~\cite{cosmic_galois_paper}. These `extended' Steinmann relations have greatly facilitated the computation of high-loop six- and seven-particle BDS-like normalized amplitudes~\cite{Caron-Huot:2016owq,Dixon:2016nkn,Drummond:2018caf}, and it was in this context that cluster adjacency was originally proposed as a generalization of the Steinmann relations. In section~\ref{sec:cluster_adjacency_implies_steinmann} we will demonstrate that the extended Steinmann relations are implied by cluster adjacency.

\subsection{The Minimally Normalized MHV Amplitude} \label{sec:minimal_subtraction}

Let us now consider the case where $n$ is a multiple of 4. Despite the fact that no BDS-like decomposition exists in these kinematics, one can still define a normalization factor that only depends on two-particle invariants. To do so, however, requires giving up dual conformal invariance~\cite{Golden:2018gtk}. While any finite amplitude resulting from such a normalization scheme has a perfectly well-defined symbol, letters involving the `infinity twistor' $\langle i\,i{+}1\,I\rangle$ will not cancel out. This is lamentable, but it is not a fatal blow to cluster adjacency: if we simply treat these letters as additional frozen variables (i.e., declare that they are adjacent to every $\mathcal{A}$-coordinate), then it can still be checked whether or not the symbol---which is expressible in terms of $\mathcal{A}$-coordinates (although of course not $\mathcal{X}$-coordinates)---obeys cluster adjacency.

Towards this end, we define a minimal normalization scheme by
\begin{align} \label{eq:min_norm_def}
{ A}_n^{\text{min}} \equiv \text{exp} \left[ \sum_{L=1}^{\infty} g^{2L} \left( - \frac{f^{(L)}(\epsilon)}{2 L^2 \epsilon^2} \sum_{i=1}^n \left(\frac{\mu^2}{-s_{i,i+1}} \right)^{L \epsilon} + C^{(L)} \right) \right],
\end{align}
where $\mu$ is the renormalization scale, and the rest of the quantities are as in~\eqref{eq:BDS_like_def}. This amounts to replacing ${ A}_n^{(1)}$ in the BDS ansatz with just its infrared-divergent part. We then define the minimally-normalized (MHV) amplitude to be
\begin{align} \label{eq:min_amp_def}
{\cal{E}}_n^{\rm min} \equiv \left. \frac{{ A}_n}{{ A}_n^{\text{min}}}\right|_{\epsilon=0} = \exp \left( R_n + \frac{\Gamma_\text{cusp}}{4} {F}_n \right).
\end{align}
As in the previous section, the fact that ${ A}_n^{\text{min}}$ depends only on two-particle Mandelstam invariants means that ${\cal{E}}_n^{\rm min}$ has the same cluster adjacency properties (and satisfies the same Steinmann relations) as the amplitude ${\cal{A}}_n$ itself. Moreover, ${\cal{E}}_n$ and ${\cal{E}}_n^{\rm min}$ can only differ by products of logarithms of two-particle Mandelstam invariants (whenever the former can be defined). We provide an explicit all-$n$ formula for the one-loop minimally-normalized MHV amplitude in appendix~\ref{sec:appendix_min_normalized}.

\section{Cluster Adjacency of MHV Amplitudes}
\label{sec:camhv}

As reviewed in the Introduction, it was conjectured in~\cite{Drummond:2017ssj} that two $\mathcal{A}$-coordinates can appear next to each other in the symbol of $\mathcal{E}_n$ if and only if they exist together in some cluster of $\Gr(4,n)$. A function satisfying this rather striking property is said to exhibit \emph{cluster adjacency}. This conjecture has been supported by an explicit analysis of amplitudes at relatively low $n$~\cite{Drummond:2017ssj,Drummond:2018dfd}. Our goal is to provide additional evidence for the cluster adjacency of $\mathcal{E}_n$ by verifying that it satisfies cluster adjacency at one and two loops for all $n$ not divisible by 4. To cover the case where $n$ is a multiple of 4, we additionally show that the minimally-normalized amplitude defined by~\eqref{eq:min_amp_def} also satisfies cluster adjacency at one and two loops for all $n$.

\subsection{One Loop} \label{sec:one_loop}

As discussed in section~\ref{sec:minimal_subtraction}, $\mathcal{E}_n$ satisfies cluster adjacency if and only if ${\cal{E}}_n^{\rm min}$ does, since they are related to each other by (products of) logarithms whose arguments involve only two-particle invariants. However, while ${\cal{E}}_n^{\rm min}$ is more general (insofar as it can be defined at all $n$), it is more transparent at one loop to see that $\mathcal{E}_n$ satisfies cluster adjacency; we thus consider this case first.

Expanding~\eqref{eq:BDS_like_normalized_def} to ${\cal O}(g^2)$, and noting that the remainder function $R_n$ is by definition zero at one-loop order, we have
\begin{align}
{\cal E}_n^{(1)} = - Y_n\,.
\end{align}
By computing $Y_n$ as described above~\eqref{eqn:ydef}, we find that the symbol of the one-loop BDS-like normalized MHV amplitude is given by
\begin{align}
        \label{eqn:oneloopsymbol}
        {\cal S}(\mathcal{E}_n^{(1)}) = \sum_{i=1}^n \left[ \sum_{j=i+3}^{n+i-3}
        u_{i,j} \otimes \langle i{-}1\,i\,i{+}1\,j \rangle - S_i(n) \otimes s_{i+1,i+2}\right],
\qquad 4\! \not\vert \  n\,,
\end{align}
where, as always, all indices are taken mod $n$, we define the cross-ratios
\begin{align}
\label{eqn:udef}
u_{i,j} = \frac{\langle i{-}1\,i\,j\,j{+}1\rangle
        \langle i\,i{+}1\,j{-}1\,j\rangle}{\langle
        i{-}1\,i\,j{-}1\,j\rangle \langle i\,i{+}1\,j\,j{+}1\rangle} \, ,
\end{align}
and
\begin{align*}
S_{i}(4k+1) &= \sum_{j=1}^{k-1}  ( s_{i+4j,i+4j+1} \ s_{i-4j+2,i-4j+3} ) - \sum_{j=1}^{k-2} ( s_{i+4j+2,i+4j+3} \ s_{i-4j,i-4j+1}) \\
&\qquad \qquad+  \frac{ s_{i-1,i} \ s_{i+3,i+4} }{ s_{i-3,i-2} \ s_{i,i+1,i+2} \ s_{i+1,i+2,i+3} \ s_{i+5,i+6} }\, , \\
S_i(4k+2) &= \sum_{j=1}^{k}  s_{i+4j,i+4j+1} - \sum_{j=1}^{k-1}  s_{i+4j+2,i+4j+3} +   \frac{s_{i+1,i+2}}{ s_{i,i+1,i+2} \ s_{i+1,i+2,i+3}}\,, \\
S_i(4k+3) &= \sum_{j=1}^{k}  ( s_{i+4j,i+4j+1} \ s_{i-4j+2,i-4j+3} )  - \sum_{j=1}^{k-1} ( s_{i+4j+2,i+4j+3} \ s_{i-4j,i-4j+1})  \\
&\qquad \qquad+ \frac{ (s_{i+1,i+2})^2 }{ s_{i-1,i} \ s_{i,i+1,i+2} \ s_{i+1,i+2,i+3} \ s_{i+3,i+4} } \,.
\end{align*}
In these formulas one may as well treat $s_{i,i+1,\dots, j}$ as shorthand for $\langle i\m 1\,i\,j\,j \p1\rangle$ since the spinor products in the denominator of~\eqref{eqn:mandelstam_translation} must all cancel out due to dual conformal symmetry.

To establish the cluster adjacency of~\eqref{eqn:oneloopsymbol}, we need to show that all the cluster ${\cal A}$-coordinates appearing in $u_{i,j}$ exist in a cluster together with $\langle i{-}1\,i\,i{+}1\,j \rangle$; all other terms involve frozen ${\cal A}$-coordinates, and automatically satisfy cluster adjacency. Thanks to the cyclic invariance of the structure of the cluster algebra (more on this below), it suffices to specialize to the $i=2$ term in~\eqref{eqn:oneloopsymbol}. Moreover it suffices to restrict to the range $4 < j < n$, since $\langle i{-}1\,i\,i{+}1\,j\rangle$ is frozen for $i=2$ and $j=4$ or $n$. The four ${\cal A}$-coordinates that appear in $u_{2,j}$ are
\begin{align}
	\label{eqn:tocheck}
	\langle 1\,2\,j{-}1\,j\rangle\,,
	\langle 1\,2\,j\,j{+}1\rangle\,,
	\langle 2\,3\,j{-}1\,j\rangle\,,
	\langle 2\,3\,j\,j{+}1\rangle \, .
\end{align}
In order to check whether each of these exists in a cluster with $\langle 1\,2\,3\,j\rangle$ we compute the relevant Sklyanin brackets, finding for $5 < j < n$
\begin{align}
\{ \log \langle 1\,2\,j{-}1\,j \rangle, \log \langle 1\,2\,3\,j \rangle \} &= 0 \, ,\\
\{ \log \langle 1\,2\,j\,j{+}1 \rangle, \log \langle 1\,2\,3\,j \rangle \} &= 0 \, ,\\
\{ \log \langle 2\,3\,j{-}1\,j \rangle, \log \langle 1\,2\,3\,j \rangle \} &= 0 \, ,\\
\{ \log \langle 2\,3\,j\,j{+}1 \rangle, \log \langle 1\,2\,3\,j \rangle \} &= 0 \, ,
\end{align}
and for the special case $j=5$
\begin{align}
\{ \log \langle 1\,2\,4\,5 \rangle, \log \langle 1\,2\,3\,5 \rangle \} &= \tfrac{1}{2} \, ,\\
\{ \log \langle 2\,3\,4\,5 \rangle, \log \langle 1\,2\,3\,5 \rangle \} &= \tfrac{1}{2} \,.
	\label{eqn:noncyclic}
\end{align}
Applying the criterion at the end of section~\ref{sec:sklyanin}, we conclude that for all $j$, $\langle 1\,2\,3\,j\rangle$ exists in a cluster with each of the four $\mathcal{A}$-coordinates displayed in~\eqref{eqn:tocheck}, and ${\cal E}_n^{(1)}$ is thereby cluster adjacent.

By cyclic invariance, the same conclusion holds for every term in the sum over $i$ in~\eqref{eqn:oneloopsymbol}. We should emphasize, however that the actual \emph{value} of the Sklyanin bracket on $\mathcal{A}$-coordinates is not cyclic invariant. For example, in contrast to~\eqref{eqn:noncyclic} we find after shifting each index by one that
\begin{align}
	\label{eqn:noncyclic2}
	\{ \log \langle 3\,4\,5\,6 \rangle, \log \langle 2\,3\,4\,6 \rangle \} = 0\,.
\end{align}
This is not a contradiction, because we do not ascribe any specific meaning to the value of the Sklyanin bracket (and indeed, note that the gauge-fixed matrix~\eqref{eq:gauge_fixed_Z} clearly breaks cyclic symmetry). We are only interested in the binary question of whether or not the Sklyanin bracket evaluates to a value in $\frac{1}{2}\mathbb{Z}$. In this respect,~\eqref{eqn:noncyclic} and~\eqref{eqn:noncyclic2} are perfectly consistent.

Let us now consider the minimally normalized amplitude ${\cal{E}}_n^{\rm min, (1)}$. We can similarly enumerate all pairs of cluster ${\cal A}$-coordinates that appear next to each other in the symbol of this amplitude, using the form of this function given in appendix~\ref{sec:appendix_min_normalized}. In the terms involving products of logarithms, it is easy to read off what coordinates will appear next to each other in the symbol: all coordinates appearing in a given logarithm will appear next to all coordinates appearing in the logarithm it multiplies. Using this observation, it can be seen that the function $f(i)$ only gives rise to pairs of coordinates that either involve two-particle invariants or the same coordinate twice; it thus satisfies cluster adjacency automatically.

The pairs of coordinates that appear in adjacent entries of the symbols of $f(i,j)$ and $g(i,j)$ can also be enumerated. (The coordinates that appear next to each other in the symbol of the dilogarithms are less transparent, but easy to calculate using standard techniques; here one must also make use of Pl\"ucker relations.) In addition to pairs of coordinates of the type found in the BDS-like normalized amplitude, coordinates taking the form $\ket{1\, 2\, j \m 1\, j}$ and $\ket{2\, 3\, j \m 1\, j}$ appear next to $\ket{1\, 2\, j\, j \p 1}$. Computing the Sklyanin bracket of these pairs, we find
\begin{align} \label{eq:sklyanin_minimal}
&\{ \log \ket{1\,2\,j \m 1\,j} , \log \ket{1\, 2\, j\, j \p 1} \} = - \tfrac{1}{2} \, ,  \\
&\{ \log \ket{2\,3\,j \m 1\,j} , \log \ket{1\, 2\, j\, j \p 1} \} =  0 \, ,
 \end{align}
for $5 < j < n$. Similar to before, there exists an exceptional case due to our gauge-fixing choice, namely
\begin{align}
\{\log \ket{1\, 2\, 4\, 5} , \log \ket{1\, 2\, 5\, 6} \} = 0
\end{align}
(we ignore further cases in which one or both coordinates reduce to a two-particle invariant). As can be seen from these Sklyanin bracket values, the one-loop minimally normalized amplitude also satisfies cluster adjacency.

The fact that additional, nontrivial pairs of ${\cal A}$-coordinates appear in adjacent symbol entries of the minimally-normalized amplitude (compared to the BDS-like normalized amplitude) merits comment. The existence of these contributions seems to contradict the fact that the conversion between these two amplitudes only involves logarithms of two-particle invariants. As required to resolve this conundrum, these additional contributions cancel out in the full sums~\eqref{eq:minimally_normalized_sum_even} and~\eqref{eq:minimally_normalized_sum_odd}.\footnote{The skeptical reader is encouraged to convince themselves of this fact using the ancillary {\sc Mathematica} notebook.} (In fact, the terms involving adjacent occurrences of $\ket{2\, 3\, j \m 1\, j}$ and $\ket{1\, 2\, j\, j \p 1}$ cancel out even in the sums~\eqref{eq:g_def} and~\eqref{eq:h_def}; we include them above to make clear that cluster adjacency is manifestly satisfied in every term.) However, this cancellation only happens at the level of the symbol, and in particular involves cancellations between terms coming from dilogarithms and products of logarithms. In general, we know of no representation of ${\cal{E}}_n^{\rm min, (1)}$ in terms of dilogarithms that satisfies cluster adjacency term by term and does away with these spurious contributions.

\subsection{Two Loops}

The symbol of the $n$-point two-loop BDS remainder function $R_n^{(2)}$ was computed in~\cite{CaronHuot:2011ky}. One can relate this to the corresponding BDS-like MHV amplitude $\mathcal{E}_n^{(2)}$ using
\begin{align}
  \label{eqn:RtoE}
 \mathcal{E}_{n}^{(2)} = R_{n}^{(2)} +\frac{1}{2}{\left(\mathcal{E}_{n}^{(1)}\right)^{2}} - 2 \zeta(2) \mathcal{E}_{n}^{(1)} \, ,
\end{align}
which follows from expanding~\eqref{eq:BDS_like_normalized_def} to ${\cal O}(g^4)$. As is well-known, the symbol alphabet entering the two-loop MHV amplitude consists of all Pl\"ucker coordinates that have at least one pair of adjacent indices, together with two more complicated types of $\mathcal{A}$-coordinates:
\begin{align}
        a_{ijk} = \langle i\,i{+}1\,\overline{j} \cap \overline{k} \rangle
        &\equiv \langle i\,j{-}1\,j\,j{+}1\rangle
        \langle i{+}1\,k{-}1\,k\,k{+}1\rangle - (j \leftrightarrow k)\,,
        \\
        b_{ijk} =
\langle i(i{-}1\,i{+}1) (j\,j{+}1)(k\,k{+}1)\rangle &\equiv
        \langle i\,i{-}1\,j\,j{+}1\rangle \langle i\,i{+}1\, k\, k{+}1\rangle
        - (j \leftrightarrow k)\,.
\label{eqn:bdef}
\end{align}
We do not endeavor to write an explicit all-$n$ formula for the symbol of $\mathcal{E}_n^{(2)}$, but instead---as at one loop---to catalog all (unordered) pairs of these coordinates that appear next to each other in the symbol of $\mathcal{E}_{n}^{(2)}$.

As before, we ignore all pairs of coordinates involving two-particle invariants and repeated coordinates. Another simplification is afforded by the parity invariance of the cluster algebra on $\Gr(4,n)$ and its underlying Poisson bracket (discussed for example in~\cite{Golden:2013xva}). Specifically, the coordinates $a_{ijk}$ get mapped to the Pl\"ucker coordinate $\langle i\,i{+}1\,j\,k\rangle$ under parity (up to frozen coordinates). As the coordinates $a_{ijk}$ only appear adjacent to symbol letters that are self-conjugate under parity, or next to Pl\"ucker coordinates whose parity conjugate is another Pl\"ucker coordinate, this means we can use parity to exchange every $\mathcal{A}$-coordinate of type $a_{ijk}$ for its parity conjugate, without increasing the complexity of the coordinate it appears next to.

At this stage we have reduced the list of adjacent coordinates that need to be considered to three basic types: $(p,p)$, $(p,b)$, and $(b,b)$, where $p$ stands for a Pl\"ucker coordinate and $b$ stands for a coordinate of the type defined in~\eqref{eqn:bdef}. Here we provide a complete classification of the $(p,p)$ pairs that appear in $\mathcal{S}(\mathcal{E}_n^{(2)})$ (in each case it is to be understood, of course, that all dihedral images of the indicated pair also appear). Using notation $(i,j) \notInt (k,l)$ that will be defined below, we have:
\begin{itemize}
	\item{$(\langle 1\,2\,3\,i\rangle, \langle 1\,2\,i\,j\rangle)$}
	\item{$(\langle 1\,2\,j\,j{+}1\rangle, \langle
		k\,k{+}1\,l\,l{+}1\rangle)$, as long as
		$(1,j) \notInt (k,l)$}
	\item{$(\langle 1\,2\,3\,j\rangle,
		\langle k{-}1\,k\,k{+}1\,l\rangle)$, as long
		as $(2,j) \notInt (k,l)$}
	\item{$(\langle 1\,2\,3\,j\rangle, \langle k\,k{+}1\,l\,l{+}1\rangle)$,
		as long as $(2,j) \notInt (k,l)$ and $(2,j) \notInt (k+1,l+1)$}
	\item{$(\langle 1\,2\,i\,i{+}1\rangle, \langle 1\,2\,j\,k\rangle)$,
		for $j \in \{i,i{+}1\}$ and $j+2 \le k \le n-2$}
	\item{$(\langle 1\,2\,i\,i{+}1\rangle, \langle 1\,2\,j\,k\rangle)$,
		for $k \in \{i,i{+}1\}$ and $4 \le j \le i+1$}
	\item{$(\langle 1\,2\,3\,j\rangle,
		\langle 2\,j\,k\,k{+}1\rangle)$,
		for $j < k \le n$}
\end{itemize}
In this classification we see a strong echo of the ``emergent planarity'' that has been observed~\cite{Arkani-Hamed:2013kca} in the structure of unitarity cuts (or equivalently, boundaries of the amplituhedron) for multi-loop amplitudes in planar sYM theory (see also~\cite{Prlina:2017tvx} for a discussion of this phenomenon in the context of symbol alphabets). Emergent planarity means the following: mark $n$ points in cyclic order around a circle, and let $(i,j)$ denote the chord connecting points $i$ and $j$. We define $(i,j) \notInt (k,l)$ to mean that the chords $(i,j)$ and $(k,l)$ do not intersect in the interior of the circle (if they touch a common vertex on the circle, they are not said to cross).

For all pairs of Pl\"ucker coordinates listed in the above catalog, the Sklyanin bracket evaluates to a half integer, supporting the conclusion that all pairs of these types exist in a cluster together.  In fact, all of these are examples of a phenomenon known as ``weak separation'' in the mathematics literature, where it has been proven~\cite{oh2015weak} that two Pl\"ucker coordinates $\langle i_1\cdots\rangle$ and $\langle j_1\cdots\rangle$ are cluster adjacent if and only if $i \notInt j$ for all pairs drawn from the sets $i \in \{i_1,\ldots\} \setminus \{j_1,\ldots\}$ and $j \in \{j_1,\ldots\} \setminus \{i_1,\ldots\}$.

The remaining pairs of $\mathcal{A}$-coordinates, of type $(p,b)$ and $(b,b)$, that appear in the symbols of two-loop MHV amplitudes are too numerous to enumerate explicitly, but with the aid of a computer algebra system to help with the bookkeeping, we have verified that the same conclusion holds for all pairs through $n=22$. This is more than enough to be confident that the result holds for all $n$, as the structure of two-loop MHV amplitudes stabilizes at much smaller $n$.

As in the previous section, we carry out the same analysis on the two-loop minimally-normalized amplitude
\begin{align}
{\cal{E}}_n^{\rm min, (2)} = R_{n}^{(2)} +\frac{1}{2}{\left(F_n \right)^{2}} +2 \zeta(2) F_n \, ,
\end{align}
which additionally covers kinematics in which $n$ is a multiple of four. We find that the Sklyanin bracket again evaluates to half-integers for all pairs of coordinates appearing in adjacent symbol entries. This concludes our verification that (appropriately normalized) MHV amplitudes in perturbative sYM theory satisfy cluster adjacency at one and two loops for all $n$.

\section{Cluster Adjacency and the Steinmann Relations} \label{sec:cluster_adjacency_implies_steinmann}

In the previous section we used the Sklyanin bracket to show that all $\mathcal{A}$-coordinates known to appear in adjacent symbol entries of individual MHV amplitudes also appear together in at least one cluster of $\Gr(4,n)$. In this section we investigate a general implication of cluster adjacency.  Specifically, using the correspondence between Mandelstam invariants and Pl\"ucker coordinates in~\eqref{eqn:mandelstam_translation}, we show that cluster adjacency implies the Steinmann relations.

As discussed in section~\ref{sec:mhv_amplitudes}, the Steinmann relations amount to the statement that certain ``Mandelstam-type'' Pl\"ucker coordinates cannot appear in adjacent symbol entries of ${\cal E}_n$ or ${\cal{E}}_n^{\rm min}$~\cite{Caron-Huot:2016owq,Dixon:2016nkn}. Using the notation introduced in the previous section, it is easy to summarize the forbidden pairs by saying that they take the form $(\langle i\,i{+}1\,j\,j{+}1\rangle, \langle k\,k{+}1\,l\,l{+}1\rangle)$, where the chords $(i,j)$ and $(k,l)$ intersect. But a special case of the weak separation principle~\cite{oh2015weak} states that
\begin{framed}
\begin{center}
\vspace{-.4cm}
Two (non-frozen) Mandelstam-type $\mathcal{A}$-coordinates $\langle i\,i{+}1\,j\,j{+}1\rangle$ \\ and $\langle k\,k{+}1\,l\,l{+}1\rangle$ are cluster adjacent if and only if $(i,j) \notInt (k, l)$.
\vspace{-.4cm}
\end{center}
\end{framed}
\noindent
Thus, as anticipated by~\cite{Drummond:2017ssj}, cluster adjacency implies the Steinmann relations for all $n$. Furthermore, cluster adjacency gives rise to no further constraints in the sector of Mandelstam-type Pl\"ucker coordinates beyond what is already implied by the Steinmann relations. Here we see again a clear hint of a connection to the emergent planarity of~\cite{Arkani-Hamed:2013kca}, which warrants further elucidation.

It is instructive to check that the Sklyanin bracket test of section~\ref{sec:sklyanin-a-test} leads to the same conclusion. Using dihedral invariance, all pairs of (non-identical) Mandelstam-type Pl\"uckers can can be mapped to the pair $(\langle 1\,2\,j\,j{+}1\rangle, \langle k\m 1 \,k \,l\,l{+}1\rangle)$ for $i$, $j$, and $k$ in one of the ranges\footnote{In fact, this mapping only uses part of the dihedral group; for any fixed $n$, we can map the third range to the first using the fact that all indices are $(\text{mod}\ n)$.}
\begin{align*}
1 < k < l < j, \qquad 2< k \le j < l, \qquad 3 < j < k < l.
\end{align*}
Except for the cases involving frozen coordinates (namely, when $j=3$ or $l=k+1$), all pairs of coordinates corresponding to the second range of indices are prohibited from appearing in adjacent symbol entries by the Steinmann relations. Conversely, none of the pairs in first or third ranges are prohibited by these relations.

We now compute the Sklyanin bracket of all of these pairs. In the first and third range of indices, these brackets evaluate to
\begin{align}
\label{eq:mandelstamgood}
&\hspace{3.92cm} \big\{ \log \ket{1\, 2\, j\, j\p1}, \log \ket{k \m 1\, k\, l\, l\p 1} \big\} = \nonumber \\
&-\frac{1}{2}\Big(H(5 \m k) + H(4 \m k) - H(3 \m k) - H(2 \m k) - H(4 \m j) + H(4 \m l) \\
&\qquad \quad - H(2 \m k \p j) - H(1\m k \p j) + H(5 \m k \m l) + H(1 \p l \m j) + H(1 \m j \p k) \Big) \nonumber \\
&\hspace{3.83cm} \text{for } ( 1 < k < l < j ) \text{ or } (3 < j < k < l) \, ,   \nonumber
\end{align}
where $H(x)$ is the discrete Heaviside function that evaluates to $1$ for all $x \ge 0$, and to 0 otherwise. In the second range of indices (again, restricting to non-frozen coordinates), we find
\begin{align}
\label{eq:mandelstambad}
\{ \log \ket{1\, 2\, j\, j\p1}, \log \ket{k \m 1\, k\, l\, l\p 1} \} &\notin \tfrac{1}{2} \mathbb{Z} \qquad \text{ for } (2< k \leq j < l) \, .
\end{align}
The results~(\ref{eq:mandelstamgood}) and~(\ref{eq:mandelstambad}) are consistent with the expectation based on weak separation and again highlight the central proposal of our paper: that cluster adjacency can be tested by explicit computation.

\section{Conclusion}

In this paper we have highlighted the Sklyanin Poisson bracket on $\Gr(4,n)$ as an efficient way to test whether any two cluster coordinates lie in a common cluster, and hence to test whether a symbol satisfies cluster adjacency. We have further demonstrated that all suitably normalized one- and two-loop MHV amplitudes in planar sYM theory pass this test, and that any symbol satisfying cluster adjacency automatically satisfies the extended Steinmann relations.

Many interesting questions remain for future work. On the mathematical side, the most urgent is of course to prove the central conjecture (in section~\ref{sec:clusterreview}) about the structure of the infinite $\Gr(4,n>7)$ cluster algebras, on which the test crucially relies. It would be very interesting if the Sklyanin bracket could be used to uncover additional structure in these still rather untamed cluster algebras; for example, any generalization of the ``weak separation'' criterion of~\cite{oh2015weak} to non-Pl\"ucker $\mathcal{A}$-coordinates could prove very useful for uncovering additional structure in multi-loop amplitudes.

On the physics side, it would be interesting to use our methods to test the cluster adjacency of other amplitudes, including tree-level and one-loop non-MHV amplitudes, some examples of which have been considered in~\cite{Drummond:2018dfd,Drummond:2018caf}. As of now, every amplitude whose symbol is known beyond one-loop order has been shown to satisfy cluster adjacency, but it would be interesting to see if further results could be bootstrapped by adopting cluster adjacency as a hypothesis (as was done for example in~\cite{Drummond:2018caf}). Two obstacles in this direction include the fact that most of the amplitudes in this theory are not expected to be expressible in terms of polylogarithmic functions (and hence, do not have symbols as conventionally defined) beyond one loop~\cite{Paulos:2012nu,CaronHuot:2012ab,Nandan:2013ip,Chicherin:2017bxc,Bourjaily:2017bsb,Bourjaily:2018ycu,Bourjaily:2018yfy}. Even for those that are (such as MHV and NMHV amplitudes), it has not yet been established that their symbols will be expressible only in terms of $\mathcal{A}$-coordinates on $\Gr(4,n)$ (see for example~\cite{Prlina:2017tvx}). It would be interesting to learn what principle generalizes cluster adjacency to these more complicated amplitudes.

Finally, we have phrased the concept of symbol-level cluster adjacency entirely in terms of $\mathcal{A}$-coordinates, following the original reference~\cite{Drummond:2017ssj}. It is worth highlighting again that the Poisson bracket on $\cal A$-coordinates is not unique. Thus, there may exist a specific choice of the skew-symmetric matrix $\Omega$ whose values encode further information. For instance, it was noted in~\cite{Drummond:2017ssj} that integrable symbols also encode information about whether or not pairs of coordinates are connected (by an arrow) when they appear together in a cluster; while this information is captured by the Sklyanin bracket on $\cal X$-coordinates, this interpretation does not extend to ${\cal A}$-coordinates. In a similar vein, it is worth investigating whether applying cluster adjacency to $\mathcal{X}$-coordinates may uncover additional structure. It was conjectured in~\cite{Golden:2018gtk} that every (projectively invariant) symbol that satisfies cluster $\mathcal{A}$-coordinate adjacency also admits a cluster $\mathcal{X}$-coordinate adjacent form. It would be interesting to find such a representation of the one- and two-loop MHV amplitudes; to this end, the $v$ and $z$ cluster variables of~\cite{Golden:2014pua} might prove useful.

\acknowledgments

We are indebted to M.~Gekhtman, A.~Goncharov, and C.~Vergu for enlightening discussions and correspondence, and to C.~Fraser and O.~G{\"u}rdo{\u g}an for useful comments on the manuscript. This work was supported in part by a Van Loo Postdoctoral Fellowship (JG); ERC Starting Grant No.~757978, a Carlsberg Postdoctoral Fellowship (CF18-0641), and a grant from the Villum Fonden (AJM); the US Department of Energy under contract {DE}-{SC}0010010 Task A (MS, AV); and by Simons Investigator Award \#376208 (AV).

\appendix

\section{The Minimally Normalized One-Loop MHV Amplitude}
\label{sec:appendix_min_normalized}

\noindent The minimally normalized amplitude was defined in Eqs.~\eqref{eq:min_norm_def} and~\eqref{eq:min_amp_def}. By construction, this function satisfies all Steinmann relations involving three- and higher-particle invariants, since only two-particle invariants appear in ${\cal A}_n^{\text{min}}$. At one loop, we have
\begin{align} \label{eq:min_norm_one_loop}
{\cal{E}}_n^{\rm min, (1)} =  F_n ,
\end{align}
where $F_n$ is two times the function defined in Eq.~(4.57) of~\cite{Bern:2005iz}. Using dilogarithm identities and changing some of the summation bounds, we can put this function in the form
\begin{align}
{\cal{E}}^{\rm min,(1)}_{n = 2k} &= \sum_{i=1}^n \bigg[ f(i) + g\left(i,\tfrac{n}{2}\right) + \sum_{j=2}^{\left \lfloor \! \! \frac{n-3}{2} \! \! \right \rfloor} h(i,j) \bigg], \label{eq:minimally_normalized_sum_even} \\
{\cal{E}}^{\rm min,(1)}_{n = 2k+1} &= \sum_{i=1}^n \bigg[ f(i) + \sum_{j=2}^{\left \lfloor \! \! \frac{n-3}{2} \! \! \right \rfloor} h(i,j) \bigg], \label{eq:minimally_normalized_sum_odd}
\end{align}
where
\begin{align}
	f(i) = & \ \frac{3}{2} \zeta(2) + \frac{1}{2} \log^2 \left(\frac{s_{i - 1, i, i + 1} }{s_{i, i + 1}} \right) - \log \left( \frac{s_{i, i + 1}}{s_{i, i + 1, i + 2}} \right) \log \left(\frac{s_{i + 1, i + 2} }{s_{i, i + 1, i + 2}} \right),
\end{align}
\begin{align} \label{eq:g_def}
	g(i,\tfrac{n}{2}) = & \ \frac12 \Li_2 \left( 1 - \frac{s_{i - 1, \dots, i + \tfrac{n}{2} - 2} \ s_{i, \dots, i + \tfrac{n}{2} - 1}}{s_{i - 1, \dots, i + \tfrac{n}{2} - 1} \ s_{i, \dots, i + \tfrac{n}{2} - 2} } \right) \nonumber \\
&\quad + \frac{1}{2} \log^2 \left(\frac{s_{i, \dots, i + \tfrac{n}{2} - 2} }{s_{i, \dots, i + \tfrac{n}{2} - 1}} \right) - \frac{1}{4} \log^2\left( \frac{s_{i - 1, \dots, i + \tfrac{n}{2} - 1}}{s_{i, \dots, i + \tfrac{n}{2} - 2} }\right)  \\
&\quad - \frac{1}{2} \log \left( \frac{s_{i - 1, \dots, i + \tfrac{n}{2} - 2} \ s_{i, \dots, i + \tfrac{n}{2} - 1}}{s_{i - 1, \dots, i + \tfrac{n}{2} - 1} \  s_{i, \dots , i + \tfrac{n}{2} - 2}} \right) \log \left(\frac{s_{i - 1, \dots, i + \tfrac{n}{2} - 1}}{s_{i, \dots, i + \tfrac{n}{2} - 2} } \right), \nonumber
\end{align}
\begin{align} \label{eq:h_def}
	h(i,j) = &\ \Li_2 \left( 1 - \frac{s_{i - 1, \dots, i + j - 1}  \ s_{i, \dots, i + j}}{s_{i - 1, \dots, i + j} \ s_{i, \dots, i + j - 1}} \right) + \log \left( \frac{s_{i, \dots, i + j} }{s_{i - 1, \dots, i + j} } \right) \log \left( \frac{s_{i, \dots, i + j} }{s_{i, \dots, i + j - 1}} \right).
\end{align}
\noindent Note that these functions do not respect dual conformal symmetry, and thus when translated into Pl\"ucker coordinates the spinor products $\langle i\,  i{+}1\, I \rangle$ do not cancel.

The utility of this form of $F_n$ is that each term in $f(i)$, $h(i,j)$, and $g(i,\tfrac{n}{2})$ separately satisfies cluster adjacency for arbitrary integer values of $i$, $j$, and $\tfrac{n}{2}$. As discussed in section~\ref{sec:one_loop}, this seems to require the introduction of spurious terms that cancel out in the overall sums~\eqref{eq:minimally_normalized_sum_even} and~\eqref{eq:minimally_normalized_sum_odd}; however, it greatly simplifies the proof that all ${\cal A}$-coordinates appearing in the adjacent symbol entries of~\eqref{eq:min_norm_one_loop} have half-integer Sklyanin bracket, allowing us to conclude that it satisfies cluster adjacency at all $n$.

\bibliographystyle{JHEP}
\bibliography{adjacency}

\end{document}